\documentclass[11pt]{article}
\pdfoutput=1

\usepackage{jheppub} 
\usepackage{amsmath,amssymb,epsfig,amsfonts}
\usepackage{epsf}
\usepackage{makeidx}
\usepackage{slashed}
\usepackage{verbatim} 
\usepackage{float}
\restylefloat{table}
\usepackage{pdflscape}
\usepackage{tikz}
\usepackage{pifont}
\usepackage{array}
\usepackage{youngtab}
\usepackage{multirow}
\usepackage{xcolor}
\usepackage{ulem}
\usepackage{cleveref}
\usepackage{tensor}
\usepackage{todonotes}
\usepackage{mathrsfs}
\usepackage{changepage}
\usepackage{anyfontsize}
\usepackage{caption}

\makeatletter

\newcommand{\be}{\begin{equation}}
\newcommand{\ee}{\end{equation}}
\newcommand{\ba}{\begin{aligned}}
\newcommand{\ea}{\end{aligned}}

\newcommand{\dd}{\mathrm{d}}
\newcommand{\me}{\mathrm{e}}
\newcommand{\ii}{\mathrm{i}}
\newcommand{\vol}{\mathrm{vol}}

\def\Im{\mathop{\mathrm{Im}}\nolimits}
\def\Re{\mathop{\mathrm{Re}}\nolimits}

\newcommand{\M}{\mathcal{M}}

\newcount\hour \newcount\minute
\hour=\time \divide \hour by 60
\minute=\time
\def\now{%
\ifnum \hour<13
  \ifnum \hour=0 \advance \hour by 12 \number\hour:\else \number\hour:\fi%
     \ifnum \minute<10 0\fi%
     \number\minute%
\ A.M.%
\else \advance \hour by -12 \number\hour:%
  \ifnum \minute<10 0\fi%
  \number\minute%
  \ P.M.%
\fi%
}

\makeatother

\begin{document}

\baselineskip=18pt  
\numberwithin{equation}{section}  
\allowdisplaybreaks  


\thispagestyle{empty}

\vspace*{0cm} 
\begin{center}
{\fontsize{18}{20}\selectfont {{\bf M2-branes on Discs and Multi-Charged Spindles}}}
 \vspace*{1cm}
 
\begin{center}

{\fontsize{12.3}{17}\selectfont Christopher Couzens$^{a ,}$\footnote{cacouzens@khu.ac.kr}, Koen Stemerdink$^{b ,}$\footnote{k.c.stemerdink@uu.nl}, 
 Damian van de Heisteeg$^{b ,}$\footnote{d.t.e.vandeheisteeg@uu.nl}
}
\end{center}
\vskip .2cm

 \vspace*{0.5cm} 
 
$^{a}$ \emph{Department of Physics and Research Institute of Basic Science,\\
  Kyung Hee University, Seoul 02447, Republic of Korea}\\
 
$^{b}$\emph{ Institute for Theoretical Physics, Utrecht University \\
 Princetonplein 5, 3584 CC Utrecht, The Netherlands}\\
  
 {\tt {}}

\vspace*{0.8cm}
\end{center}

 \renewcommand{\thefootnote}{\arabic{footnote}}

\begin{adjustwidth}{0.26in}{0.26in}

\begin{center} {\bf Abstract } \end{center}

\vspace{0.32cm}

\noindent

We study supersymmetric AdS$_2\times Y_9$ solutions of 11d supergravity where $Y_9$ is an $S^7$ fibration over a Riemann surface equipped with a metric of non-constant curvature. We consider two classes of Riemann surface: the first is a spindle and the second is a topological disc. These solutions are interpreted as the near-horizon limit of M2 branes wrapped on the Riemann surface and describe the near-horizon of a 4d black hole. In the case of the topological disc there are additional flavour M2 branes smeared on a five-sphere embedded in the transverse $S^7$. We perform a full global analysis of both classes of solutions, both from a 4d and an 11d viewpoint. Finally we compute the two-dimensional Newton's constant from which we obtain a prediction for the entropy of the black hole.

\end{adjustwidth}
\noindent

\newpage


\tableofcontents
\printindex
\setcounter{footnote}{0}
\newpage

\section{Introduction}

One of the best tools for constructing field theories in lower dimensions is to compactify a higher dimensional theory on a compact manifold. If the parent theory has a brane construction one can view the compactified theory as arising from wrapping the branes on certain cycles. Following the seminal work in \cite{Maldacena:2000mw}, one may probe these constructions holographically by constructing interpolating flows across dimensions: that is AdS$_d$  geometries flowing to AdS$_{d-p}\times \Sigma_p$ geometries. The canonical method of preserving supersymmetry in these constructions was to perform a so called \emph{topological twist} of the theory \cite{Witten:1988ze}. The smoking gun of a topological twist, is a Killing spinor which is constant on the compactification manifold. The twist is engineered so that background R-symmetry gauge fields cancel off the spin connection such that the spinor is constant.

More recently a different realisation of preserving supersymmetry has been given in \cite{Ferrero:2020laf} for compactifications of D3 branes on a \emph{spindle}. A spindle is topologically a sphere admitting orbifold singularities at both poles, characterised by relatively prime integers $n_{\pm}$ labelling the deficit angles $2\pi (1-n_{\pm}^{-1})$. It transpires that the Killing spinors are sections of non-trivial bundles of the spindle, and thus not constant. Consequently, supersymmetry is not preserved via a topological twist. Spindle constructions have since been extended to setups involving; multi-charge spindles of D3 branes \cite{Hosseini:2021fge,Boido:2021szx},  (rotating) M2 branes \cite{Ferrero:2020twa,Cassani:2021dwa} and (multi-charge) M5 branes \cite{Boido:2021szx,Ferrero:2021wvk}.

Independently, in \cite{Bah:2021mzw,Bah:2021hei} solutions containing a topological disc and preserving supersymmetry in a similar manner have been constructed. These are dual to Argyres--Douglas theories arising from compactifying M5-branes on a sphere with two punctures, one regular and one irregular. Another interesting feature of these solutions is the presence of smeared M5-branes lying along the boundary of the disc. This construction was extended to D3-branes wrapped on topological discs in \cite{Couzens:2021tnv,Suh:2021ifj} where in the former reference it was noted that the disc can be obtained from a different global completion of the multi-charge spindle solutions of \cite{Boido:2021szx}. Later work has extended this to D4-D8 branes \cite{Suh:2021aik} wrapped on discs.  

In this paper we will consider multi-charge AdS$_2\times \Sigma$ solutions in 4d $\mathcal{N}=2$ U$(1)^4$ STU supergravity. The solutions we consider can be uplifted to 11d supergravity on $S^7$ and their local forms were originally found in this 11d guise in \cite{Gauntlett:2006ns}. We will perform a full global analysis of these solutions from both 4d and 11d, in a similar manner to the global regularity performed in \cite{Ferrero:2020twa} and \cite{Bah:2021hei}. We will show that different global completions of the solution give rise to two classes of solutions, one where $\Sigma$ is a spindle and another where $\Sigma$ is a topological disc. We appropriately quantise the fluxes of the four abelian gauge fields so that the fibration 
\be
\text{S}^7 \hookrightarrow Y_9\rightarrow \Sigma\, ,
\ee
is well-defined before considering the full 11d solutions. We show in the spindle class of solution that the internal metric $Y_9$ is smooth. For the disc solution we show that the solution takes the form of a $S^5\times S^1_z\times S^1_{\phi_4}$ fibration over a rectangle. On the boundary of the rectangle we show that the metric degenerates smoothly except at two of the four corners with the singularities associated to the presence of smeared M2 branes and a monopole respectively. 

This paper is organised as follows. In section \ref{sec:4d} we study the AdS$_2\times \Sigma$ solutions in 4d $\mathcal{N}=2$ U$(1)^4$ gauged STU supergravity. We study the regime in which the solutions are well-defined and impose appropriate quantisation conditions in preparation for uplifting the solutions to 11d in section \ref{sec:11d}. We show explicitly that supersymmetry is not realised by a topological twist by computing the Euler character of $\Sigma$ and comparing this with the sum of the charges of the gauge fields. In section \ref{sec:11d} we consider the uplifted solutions on an $S^7$, presenting the solutions in both the canonical AdS$_2\times Y_9$ form of \cite{Kim:2006qu} and in the canonical uplift form. In the spindle class we show that the solutions are regular. For the disc class we show that the singularities in 4d may be interpreted as a smeared M2-branes and the existence of a monopole. Finally we quantise the flux and compute the on-shell Newton's constant, giving a prediction for the entropy of a putative asymptotically AdS$_4$ black hole with horizon $\Sigma$. 
We conclude in section \ref{sec:conclusion} and relegate some technical material on smeared M2-branes to appendix \ref{app:smeared}.

{\bf Note added:} Whilst we were preparing the manuscript for publication \cite{Suh:2021hef} appeared on the arXiv, which has some overlap with the topological disc solutions presented here. We find that some of our regularity analysis differs with that conducted there. In addition, \cite{Ferrero:2021ovq} appeared which has overlap with the spindle solutions presented here.


\section{4d Black Hole Near-Horizons}\label{sec:4d}

In this section we will study a family of supersymmetric AdS$_2$ solutions in 4d U$(1)^4$ gauged supergravity which may be uplifted to solutions of 11d supergravity on an $S^7$. The local form of the 11d solutions we study was originally found in \cite{Gauntlett:2006ns} using dualities between multi-charge superstar solutions. We will study in detail the 4d AdS$_2\times \Sigma$ solutions in this section. One finds that there are two distinct ways of extending the local solutions globally, distinguished by the different properties of $\Sigma$. The first class gives rise to a spindle, $\mathbb{WCP}^{1}_{[n_-,n_+]}$. Whilst in the second class of solution, $\Sigma$ is a topological disc and is (naively) singular. As we will see in section \ref{sec:11d}, this singularity arises due to the presence of smeared M2-branes in the full 11d solution and is therefore physical. 


\subsection{Multi-charge solutions of 4d U$(1)^4$ gauged supergravity}

The solutions we will study arise in 4d $\mathcal{N}=2$ U$(1)^4$ gauged supergravity which is a consistent truncation of $\mathcal{N}=8$ SO$(8)$ gauged supergravity. The action, following the conventions in \cite{Azizi:2016noi}, with which we may uplift solutions to 11d on $S^7$, is\footnote{For ease of notation we have set the coupling constants for each of the gauge fields to 1 without loss of generality.}
\begin{align}
S=\frac{1}{16 \pi G_{(4)}} \int \bigg( R-\frac{1}{2} \sum_{I=1}^{4} (X^{(I)})^{-2} \big(\dd X^{(I)}\big)^2 + \sum_{I<J} X^{(I)}X^{(J)}-\frac{1}{2} \sum_{I} \big(X^{(I)}\big)^{-2} \big|F^{I}\big|^2\bigg)\dd \vol_4\, ,\label{eq:action}
\end{align}
subject to $X^{(1)}X^{(2)}X^{(3)}X^{(4)}=1$.
One may obtain the above action from the general 4d $\mathcal{N}=2$ supergravity action by using the pre-potential\footnote{See for example \cite{Lauria:2020rhc}.}
\be
F=-\ii \sqrt{X^{(1)}X^{(2)}X^{(3)}X^{(4)}}\,,
\ee
see appendix \ref{app:spinors} for further details. There are three further consistent truncations of the theory of interest to us named T$^3$, $X^0 X^1$ and Einstein--Maxwell and are specified by reducing the independent scalars and gauge fields as given in table \ref{table:truncation}. 
\begin{table}[h]
\begin{center}
\begin{tabular}{|c|c|c|}
\hline
Theory & Scalars & Gauge Fields\\
\hline\hline
T$^3$  &$X^{(1)}=X^{(2)}=X^{(3)}$& $A_{1}=A_{2}=A_{3}$\\
\hline
$X^0 X^1$ & $X^{(1)}=X^{(3)}\, ,\, \, X^{(2)}=X^{(4)}$ & $A_1=A_3\, ,\, \, A_2=A_4$\\
\hline
Einstein--Maxwell & $X^{(1)}=X^{(2)}=X^{(3)}=X^{(4)}=1$ & $A_1=A_2=A_3=A_4$\\
\hline
\end{tabular}

\end{center}

\caption{The three consistent truncations of 4d $\mathcal{N}=2$ U$(1)^4$ gauged supergravity. We will largely ignore the Einstein--Maxwell truncation since it has appeared previously in the literature, and can be viewed as a special case of the other two truncations.}
\label{table:truncation}
\end{table}

The local AdS$_2$ solutions that we will consider here are obtained from truncating the 11d solutions in \cite{Gauntlett:2006ns} to 4d. Truncating the aforementioned solution of the seven-sphere, one obtains an AdS$_2$ solution to 4d U$(1)^4$ gauged supergravity. The bosonic sector is given by
\begin{align}
\dd s^2_4&= \sqrt{P(w)} \bigg( \dd s^2 (\text{AdS}_2) +\dd s^2(\Sigma)\bigg)\, ,\label{eq:4dmetric}\\
\dd s^2(\Sigma)&=\frac{f(w)}{P(w)} \dd z^2 + f(w)^{-1} \dd w^2\, ,\label{eq:Sigma}\\
A_{I}&=-\frac{w}{2(w-q_{I})} \dd z\, ,\label{eq:gauge}\\
X^{(I)}&= \frac{P(w)^{1/4}}{w-q_{I}}\, ,\label{eq:scalars}
\end{align}
where the polynomials $f(w)$ and $P(w)$ are 
\be
P(w)=\prod_{I=1}^{4} (w-q_{I})\, ,\qquad f(w)= P(w)-w^2\, .\label{Pandf}
\ee
The solution is specified by four constants, $q_{I}$, which are in principle independent. In the following section we will extend this local solution to a well-defined global solution which requires constraints on the parameters to be imposed. We will show that the topology of $\Sigma$ depends on the choice of parameters $q_{I}$ giving rise to two distinct classes of solutions.


\subsection{Global Analysis}
In order to extend the local solution \eqref{eq:4dmetric}-\eqref{eq:scalars} to a globally well defined one, one must impose a number of additional constraints. Firstly, we must require that the metric is both real and has the correct signature. This implies that both functions $f(w)$ and $P(w)$ must be positive definite. Moreover, we must fix the domain of the coordinate $w$ so that $\Sigma$ is a compact space. To do this we identify two zeroes of the function $f(w)$ between which both $f(w)$ and $P(w)$ are strictly positive. 
Since $f(w)$ is a quartic polynomial it admits four roots. Clearly we need at least two real roots, in fact we can immediately rule out $f(w)$ admitting only two real roots. This follows since $f(w)$ tends to infinity as $w\rightarrow \pm \infty$, and therefore the only domain where it is positive in this case is between the larger root and $\infty$ or $-\infty$ and the smaller root, consequently the domain is non-compact and therefore also $\Sigma$. We conclude that we must require four real roots. 

Let us denote the roots by $w_{I}$, $I=1,\ldots,4$, with the labels chosen such that $w_{1}\leq w_{2}<w_3\leq w_4$. The domain of $w$ is then $[w_2,w_{3}]$, inside which, by construction, both $P(w)$ and $f(w)$ are non-negative. Note that we do not allow $w_2=w_3$ as this would not give a finite size domain for $w$, and therefore the geometry is not well-defined. Depending on the choice of $q_I$ we will find two classes of solutions characterised by the behaviour of the metric at the end-point $w_2$. When $w_2 \neq 0$ we find that $\Sigma$ is a spindle, whilst for $w_2=0$ we encounter a topological disc. The main distinctions between these two classes have been summarised in table \ref{table:roots}.

\begin{table}[h!]
\centering
\renewcommand*{\arraystretch}{1.2}
\begin{tabular}{| c || c | c | }
\hline
&  Spindle & Topological disc \\ \hline \hline
$f(w_2)/P(w_2) $ & 0 & 1 \\ \hline
$f(w_3)/P(w_3) $ & 0 & 0 \\ \hline
parameters & \begin{minipage}{0.1\textwidth}\vspace*{0.17cm} \centering
$q_I < w_2$\\
$q_I \neq 0$ \vspace*{0.15cm}
\end{minipage}&  \begin{minipage}{0.1\textwidth}\vspace*{0.17cm} \centering
$q_i <0$ \\
$q_4 = 0 $\vspace*{0.15cm}
\end{minipage} \\ \hline
\end{tabular}
\caption{\label{table:roots} Summary of the main properties that distinguish the spindle and topological disc solutions. The ratio $f(w)/P(w)$ indicates whether the circle $S^1_z$ degenerates or remains of finite size at the endpoints of the domain $[w_2, w_3]$. 
Schematic graphs of the functions $f(w)$ and $P(w)$  for the two respective classes of solutions have also been included in figures \ref{fig:spindleplot} and \ref{fig:topdiscplot}. 
}
\end{table}

\begin{figure}[h!]
\centering
  \includegraphics[width=0.6\linewidth]{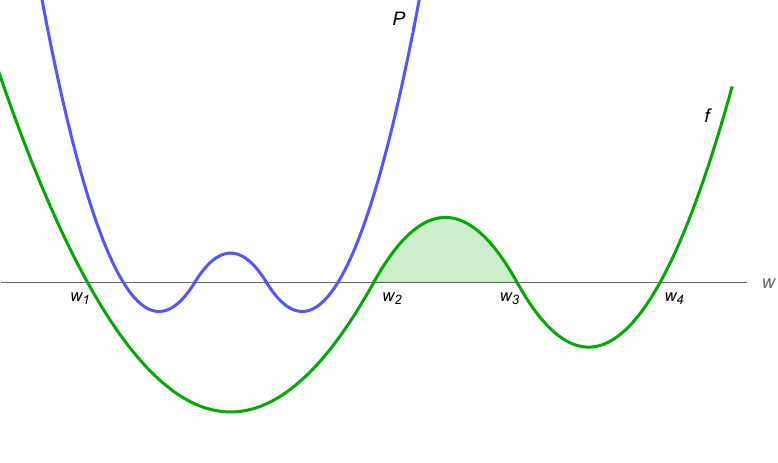}
  \captionsetup{width=.75\linewidth}
  \caption{\textit{A schematic plot showing the root structure of $f$ and $P$. Here the shaded green region indicates the domain of $w$ where the metric is defined.}}
  \label{fig:spindleplot}
\end{figure}

Before we study the metric around such end-points, let us look at the constraints for well-defined scalars. Since they are related to dilatons they must be non-negative.\footnote{As we will see later, it is in fact consistent to allow the dilatons to vanish at an end-point of the interval. The apparent singularity due to vanishing dilatons will be interpreted as the presence of a smeared M2 brane and is thus of physical nature.} This requires $w-q_{I}\geq 0$ on the domain of $w$. We may reduce this condition to $w_2\geq \text{max}\{q_I\}$ which, if we label the $q_I$ in ascending order, simply becomes $w_2\geq q_4$. Below we will study the constraints on the parameters $q_{I}$ such that the function $f(w)$ and the scalars have the properties discussed above. Schematically, all the constraints discussed above are equivalent to finding constraints on the $q_{I}$ such that the functions $f(w)$ and $P(w)$ take the schematic form given in figure \ref{fig:spindleplot}.

Before moving on to studying the constraints on the parameters $q_I$ let us now look at the charges of the solution. The solution admits four magnetic charges. Using \eqref{eq:gauge}, and parametrising our temporary ignorance of the period of the $z$ coordinate by denoting it by $\Delta z$, we find the magnetic charges
\be
Q_{I}= \frac{1}{2\pi} \int_{\Sigma} \dd A_I= \frac{q_{I}(w_3-w_2)}{2(w_3-q_{I}) (w_2-q_I)}\frac{\Delta z}{2\pi}\, .\label{eq:QIdef}
\ee
Since the charge depends explicitly on the period $\Delta z$ we will postpone quantising the fluxes for the moment. However we may still compute the sum of the charges
\be
\sum_{I=1}^{4} Q_I= \frac{\Delta z}{4\pi}\bigg(\text{sgn}(w_3)\frac{|f'(w_3)|}{|w_3|}+\text{sgn}(w_2)\frac{|f'(w_2)|}{|w_2|}\bigg)\, .\label{eq:rootsum}
\ee
Finally we can also compute the Euler character of the space\footnote{We assume that there is no boundary to $\Sigma$ here. As we will see later we must amend this, however it turns out that the boundary contribution is trivial.}
\begin{align}
\chi(\Sigma)&= \frac{1}{4\pi}\int_{\Sigma} R \dd \vol_{\Sigma}=  -\frac{w^2 f'(w)-2 w f(w)}{2 P(w)^{3/2}}\frac{\Delta z}{2\pi}\bigg|_{w=w_2}^{w=w_3}\nonumber\\
&=\frac{\Delta z}{4\pi}\bigg(\frac{|f'(w_3)|}{|w_3|}+\frac{|f'(w_2)|}{|w_2|}\bigg)\, ,\label{eq:Euler}
\end{align}
where, in the last line, we have used that $f'(w_2)>0>f'(w_3)$. Note in the two results above we have implicitly assumed that there is not a root at $0$. As we will see later this is a special point in the parameter space of the solution and requires a separate treatment. Note that the sum of the charges and Euler character can only be the same if the roots are both of the same sign, in this case we would have a solution of type `topological topological twist'.\footnote{Solutions of this form will be addressed in \cite{Couzens:2021cpk}.} 
 For roots with different sign, which we will study here, these two quantities are not equal and therefore the solution is not dual to a CFT which has been (topologically) topologically twisted.


\subsubsection{Spindle}

We now want to study how the metric degenerates around an end-point of the domain of $w$. First let us assume that the roots $w_2$ and $w_3$ are single roots\footnote{For a double root the metric around the root looks locally like hyperbolic space, see for example \cite{Gnecchi:2013mja}.} and non-zero\footnote{We will study the case where one of the roots is zero later in section \ref{sec:disc}, since this corresponds to the class of topological disc solutions rather than spindle solutions.}. Around such an end-point the metric on $\Sigma$ becomes
\begin{align}
\dd s^2(\Sigma)= \frac{1}{f'(w_*)(w-w_*)} \dd w^2 +\frac{f'(w_*)(w-w_*)}{w_*^2}\dd z^2 = \frac{4}{|f'(w_*)|} \bigg (\dd R^2 + \frac{|f'(w_*)|^2}{4 w_*^2} R^2 \dd z^2\bigg)\, ,\label{metricpole}
\end{align}
where we changed coordinates to $R^2= \pm(w-w_*)$ in the last line, taking the plus sign for the expansion around $w_2$ and the minus sign for $w_3$. Note that $f'(w_2)>0>f'(w_3)$ and the sign introduced in the definition of the new radial coordinate has been absorbed by introducing the norm. From \eqref{metricpole} we can see that the metric around the end-point looks locally like that of $\mathbb{R}^2$ if $z$ has period
\be
\frac{\Delta z}{2\pi}= \frac{2 |w_*|}{|f'(w_*)|}\, .\label{eq:periodspindlegen}
\ee
Since we have the same type of degeneration at both poles the space takes the form of a topological sphere. For this to be a round sphere we must avoid conical singularities at both poles, i.e. that the period given in \eqref{eq:periodspindlegen} is the same at both end-points
\be
\bigg|\frac{f'(w_2)}{w_2}\bigg|=\bigg|\frac{f'(w_3)}{w_3}\bigg|\, .
\ee
Using the explicit form of $f(w)$ this is equivalent to the roots satisfying
\be
|w_3|(w_2-w_1)(w_4-w_2)=|w_2|(w_3-w_1)(w_4-w_3)\, .
\ee
For the two roots having the same sign the condition reduces to\footnote{We drop any solutions which sets roots equal or to 0.}
\be
w_1w_4=w_2 w_3\, ,\label{no spindle condition}
\ee
whilst for opposite sign we have
\be
w_4=\frac{w_2 w_3 (2 w_1-w_2-w_3)}{w_1 w_2+w_1 w_3-2 w_2 w_3}\, .\label{no spindle condition2}
\ee
Since the roots of the quartic are particularly unwieldy, in order to study the solutions analytically we will consider the solution in the truncated theories. 


\paragraph{$X^0 X^1$ truncation.}

First consider the solution in the $X^0 X^1$ truncation. To truncate \eqref{eq:4dmetric}-\eqref{eq:scalars} to a solution of the $X^0 X^1$ theory we must set $X^{(1)}=X^{(2)}$, $X^{(3)}=X^{(4)}$ and $A_{1}=A_{2}$, $A_{3}=A_{4}$ as in table \ref{table:truncation}. This is equivalent to setting the constants $q_I$ to satisfy $q_1=q_2$ and $q_3=q_4$. 
In this truncation the function $f(w)$ takes the simplified form
\be
f(w)=(w-q_1)^2(w-q_3)^2-w^2\, .
\ee
Both $q_1$ and $q_3$ are non-zero, since a constant equal to zero would automatically imply a common root between $f(w)$ and $P(w)$, which we will study in the section \ref{sec:disc}. To proceed, it is useful to define
\be
s=q_1+q_3\, ,\qquad p=4 q_1 q_3\, ,
\ee
which allows us to write the four roots of $f(w)$ in the compact form
\begin{align}\label{rootsspindleST}
\frac{1}{2}\Big(s+1 \pm \sqrt{(s+1)^2-p}\Big)\,,\qquad \frac{1}{2}\Big(s-1\pm \sqrt{(s-1)^2 -p}\Big)\, .
\end{align}
Note that we have not assigned these roots the names $w_I$ with $I=1,\ldots,4$ yet, because we must still determine their order. There are two regimes where all the roots are real, distinguished by the sign of $s$. We find these regimes to be
\begin{align}\label{conditionsrealroots}
s\geq0\, ,\quad  p\leq (1-s)^2\qquad \text{or}\qquad s\leq 0 \, ,\quad p\leq (1+s)^2\, .
\end{align}
It turns out that there is only one ordering of the roots which is consistent with the positive scalar condition, $w_2-q_3>0$:
\begin{equation}\label{rootsspindleSTws}
\begin{aligned}
&w_1=\frac{1}{2}\Big(s-1- \sqrt{(s-1)^2 -p}\Big)\,,\qquad &w_2=\frac{1}{2}\Big(s-1+ \sqrt{(s-1)^2 -p}\Big)\,,\\[3pt]
&w_3=\frac{1}{2}\Big(s+1 - \sqrt{(s+1)^2-p}\Big)\,,\qquad &w_4=\frac{1}{2}\Big(s+1 + \sqrt{(s+1)^2-p}\Big)\,, 
\end{aligned}
\end{equation}
along with the additional constraints, either
\begin{align}
-1<s\leq-\tfrac{1}{2}\, ,\quad  0<p< (1+s)^2\, ,\qquad \text{or}\qquad -\tfrac{1}{2}<s< 0 \, ,\quad 0<p\leq s^2\, ,
\end{align}
which are a further restriction of the conditions in \eqref{conditionsrealroots}.
In terms of $q_1$ and $q_3$ ($q_1\leq q_3$) these conditions reduce to
\begin{equation}\label{conditionscharges}
-\tfrac{1}{4}<q_3<0 \,,\quad -\big(1-\sqrt{-q_3}\big)^2 < q_1 \leq q_3 \,.
\end{equation}
One may naturally wonder if these regimes are compatible with a spherical horizon, i.e. one where the conical singularities can be removed. It is simple to show that this is not possible and one always obtains a spindle. Plugging the roots \eqref{rootsspindleSTws} into either of \eqref{no spindle condition} or \eqref{no spindle condition2}, we find that the only possibilities to satisfy either of them are to set one of $q_1$ or $q_3$ to zero or $q_1=-q_3$. Both of these options are incompatible with the region in \eqref{conditionscharges}. The case where either $q_1$ or $q_3$ vanish needs a more careful treatment since we introduce a different degeneration of the solution as we will see in the next section. 

We conclude that it is not possible to find a spherical horizon in the $X^0 X^1$ truncation. The solutions we have found here are spindles and admit conical singularities. Following \cite{Ferrero:2020twa}, instead of making a single choice for the period $\Delta z$ we impose
\be
\Delta z=\frac{4\pi |w_2|}{n_{-}|f'(w_2)|}=\frac{4\pi |w_3|}{n_{+}|f'(w_3)|}\, ,\label{eq:spindleperiod}
\ee
which exhibits the space as the spindle $\Sigma=\mathbb{WCP}^{1}_{[n_-,n_+]}$, with conical deficit angles $2\pi (1-n_{\pm}^{-1})$ at the two poles. 

We can now return to the quantisation of the magnetic charges. Following \cite{Ferrero:2020twa} the correct quantisation condition to impose is 
\be
Q_{I}=\frac{1}{2\pi} \int_{\Sigma}F_I=\frac{p_I}{n_- n_+}\quad \text{with }\, \, p_I\in \mathbb{Z}\, .\label{eq:spindleps}
\ee
As explained in \cite{Ferrero:2020twa} with this quantisation, and for charges $p_I$ coprime to both $n_{\pm}$, this gives rise to a well-defined and smooth orbifold circle fibration of the seven-sphere over the spindle
\be
\text{S}^7 \hookrightarrow Y_9 \rightarrow \mathbb{WCP}^{1}_{[n_-,n_+]}\, .
\ee
The twist parameters of the fibration are $p_I$ and is such that it leads to a compact space $Y_9$. From \eqref{eq:QIdef} and \eqref{eq:spindleps} we find that the twist parameters are given by
\be
p_{I}= n_{-}n_{+} \frac{q_I(w_3-w_2)}{2(w_3-q_I)(w_2-q_I)}\frac{\Delta z}{2\pi}\, .
\ee
Inserting the expression for the period, \eqref{eq:spindleperiod} into the sum of the roots \eqref{eq:rootsum} and the Euler character \eqref{eq:Euler} we find
\begin{align}
\sum_{I=1}^{4}Q_{I}&=\frac{1}{n_+}-\frac{1}{n_{-}}\, ,\label{eq:spindleQsum}\\
\chi(\Sigma)&= \frac{1}{n_+}+\frac{1}{n_-}\, ,
\end{align}
confirming that this solution does not involve the usual topological twist.



\subsubsection{Topological disc}\label{sec:disc}
\begin{figure}[h!]
\centering
  \includegraphics[width=0.6\linewidth]{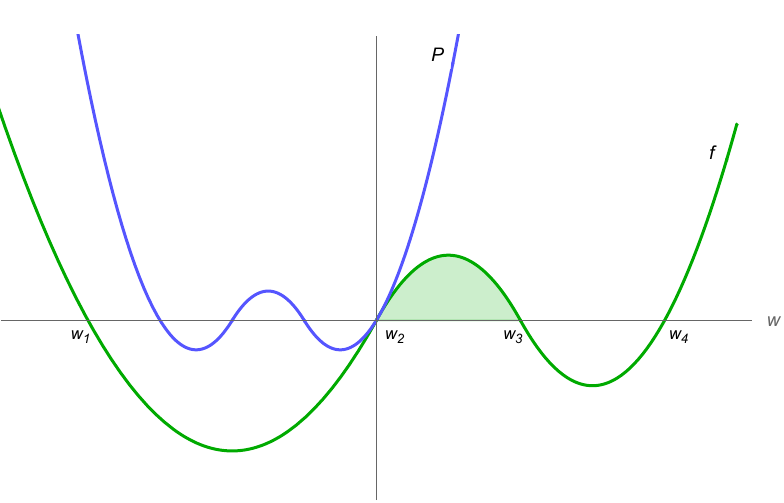}
  \captionsetup{width=.75\linewidth}
  \caption{\textit{A schematic plot showing the root structure of $f(w)$ and $P(w)$ such that we find a topological disc. As before the shaded green region indicates the domain of $w$ where the metric is defined.}}
  \label{fig:topdiscplot}
\end{figure}

Let us now consider the second way of obtaining a degeneration of the surface: a common root between $P(w)$ and $f(w)$.
We can see immediately from the form of $f(w)$, namely $f(w)=P(w)-w^2$, that if $f(w)$ and $P(w)$ have a common root, then this root must be located at $w=0$. Since the roots of $P(w)$ are given by the parameters $q_I$, this means that (at least) one of the charges must be zero. In order to find a topological disc as the black hole horizon, we want the zero root of $f(w)$ to be a boundary of the domain of $w$, i.e. it should be either $w_2$ or $w_3$ in the notation above. Recall that for the scalars to be positive we need to impose $w_2\geq\text{max}\{{q_I}\}$, which for at least one $q_I=0$ implies $w_2\geq0$. It is clear that the only consistent specification of the roots is to set $w_2=0$, and $w_3>0$ with $q_4=0$ (all other $q$'s are then $\leq0$). It follows that in order for the solution to have a root at $0$ we must require the second root of $f(w)$ and the fourth root of $P(w)$ to be equal to zero. A schematic plot of this setup is given in figure \ref{fig:topdiscplot}.

Now consider the possibility of a double root at zero. In order for this to work we have to set either $w_1$ or $w_3$ equal to zero. We can see immediately that the option $w_3=0$ is not allowed since a finite domain requires $w_3>w_2=0$. This leaves $w_1=0$. We can see from figure \ref{fig:topdiscplot} that this option would set all four $q_I$ to zero by requiring $f(w)\leq P(w)$ everywhere. Consequently in this case we would simply have $P(w)=w^4$ and $f(w)=P(w)-w^2 = w^2(w-1)(w+1)$. We see that this expression for $f(w)$ has one negative and one positive root, which is in contradiction with our choice that $w_1=w_2=0$. We conclude that the only setup with a single root at zero, i.e.~setups with a root structure as in figure \ref{fig:topdiscplot}, are the relevant ones to consider, all others do not give rise to well-defined solutions.


\paragraph{$\boldsymbol{\text{T}^3}$ truncation.}

We saw above that we must fix only one root of $f(w)$ to be zero. Consequently, this completion of the space is not possible in either the Einstein--Maxwell theory nor the $X^0 X^1$ truncation. In order to analyse this setup somewhat analytically we study this completion in the $T^3$ truncation, fixing the three non-zero charges equal, i.e. $q_1=q_2=q_3\equiv q<0$. This essentially pushes the three negative roots of $P(w)$ in figure \ref{fig:topdiscplot} together into a triple root. In this case the functions simplify to
\begin{equation}
\begin{aligned}
P(w) &\,=\, w\,(w-q)^3 \,,\\
f(w) &\,=\, w\,\big((w-q)^3 - w\big) \,.
\end{aligned}
\end{equation}
We now require the cubic polynomial $f(w)/w=(w-q)^3 - w$ to have one negative and two positive roots, so that the root equal to zero is indeed $w_2$.

It is straightforward to compute the discriminant of the polynomial $f(w)/w$, and is given by $4-27q^2$. A cubic polynomial has three real roots if its discriminant is non-negative, which gives us the constraint $-\tfrac{2}{3\sqrt{3}}\leq q\leq \tfrac{2}{3\sqrt{3}}$ for the allowed values of $q$. We have already imposed that $q$ must be negative for the positivity of the scalars, and so we restrict to the range $-\tfrac{2}{3\sqrt{3}}\leq q<0$.  

\begin{figure}[h!]
\centering
  \includegraphics[width=0.6\linewidth]{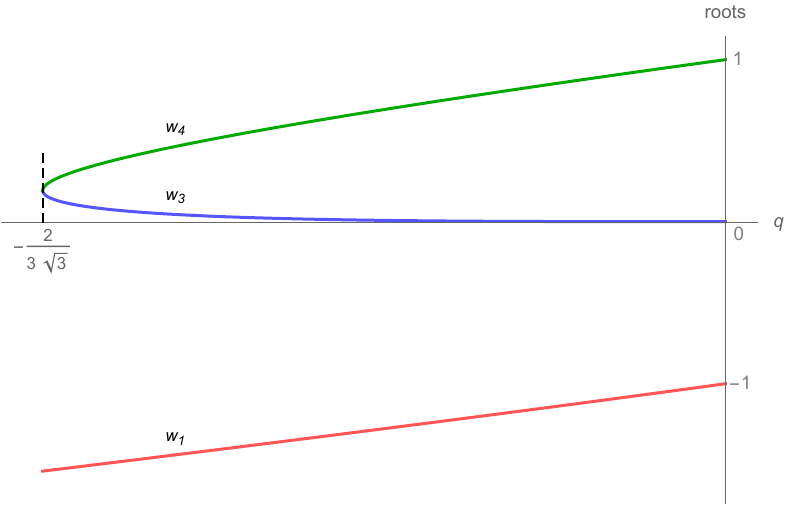}
  \captionsetup{width=.7\linewidth}
  \caption{\textit{The values of the three roots of $f(w)/w$ as a function of $q$ in the $T^3$ truncation. The root $w_3$ approaches zero as $q\rightarrow0$ but is positive for all values in the range $q\in\big[\!-\tfrac{2}{3\sqrt{3}},0\big)$.}}
  \label{fig:topdiscplotroots}
\end{figure}

The equation $f(w)/w=0$ can be solved analytically, but since the results are surprisingly bulky for such a simple polynomial equation, we will not present the explicit values here. Instead, we plot the three roots in figure \ref{fig:topdiscplotroots} to show that indeed one is negative and two are positive as required. We see that for all values of $q$ in the range $\big[\!-\tfrac{2}{3\sqrt{3}},0\big)$ we indeed obtain one negative and two positive roots, therefore for all of these values we have a solution with a horizon admitting a completion with a root at 0. 

We now want to determine the global form of the metric. At $w_3$ we may use our results in the previous section to determine that the metric on $\Sigma$ degenerates as the orbifold $\mathbb{R}^2/\mathbb{Z}_k$ if the period of $z$ is fixed to be
\be
\Delta z= \frac{4\pi w_3}{k |f'(w_3)|}\, .\label{Deltazdisc}
\ee
Consider now the degeneration at $w=0$. Note that we must consider the full 4d-solution now since the overall warp factor $P(w)$ vanishes here. Around $w=0$ the metric takes the form
\be
\dd s^2_4 =\sqrt{|q|^3w}\Big(\dd s^2 (\text{AdS}_2) +\dd z^2+\frac{1}{w |q|^3} \dd w^2\Big)\, .
\ee
This is conformal to the direct product of AdS$_2$ with a cylinder, however note that the conformal factor vanishes in this limit and the metric has a singularity at $w=0$. To add to the apparent misery, in this limit three of the four $X^{(I)}$'s vanish despite being dilatons. Fortunately all is not lost: this singularity is physical! Since this is best seen from the 11d uplifted solution we will postpone this discussion until section \ref{sec:11ddisc} and proceed unabated. We now want to ascertain what space $\Sigma$ is. We see that at $w=w_3$ the space looks like the orbifold $\mathbb{R}^2/\mathbb{Z}_k$ whilst at $w=0$ we have a circle which does not contract. This is describing a topological disc with an orbifold singularity at the centre. Consider the Euler characteristic of $\Sigma$. We find\footnote{Note that since $\Sigma$ has a boundary one must be slightly more careful with this computation than in the spindle case. The Gauss--Bonnet theorem contains a contribution from the boundary
\begin{equation*}
\chi(\Sigma)=\frac{1}{4\pi}\int_{\Sigma} R\dd \vol(\Sigma)+\frac{1}{2\pi} \int_{\partial\Sigma} \kappa \dd\vol(\partial \Sigma)\, ,
\end{equation*}
which we neglected earlier. However in the present case the geodesic curvature $\kappa$ vanishes and therefore the boundary does not contribute to the Euler characteristic.
}
\be
\chi(\Sigma)=\frac{1}{k}\, .
\ee
Recall that for an orbifold $O$, the Euler characteristic is given by $\chi(O)=\frac{1}{d}\chi(M)$ where $M$ is a compact oriented manifold providing a finite covering of $O$ of degree $d$. We therefore see that this is precisely the expected result for a $\mathbb{Z}_k$ orbifold of a disc. Next let us examine the sum of the magnetic charges of the gauge fields. Since we set $q_4=0$ we see that $A_4$ is now pure gauge, whilst the remaining three are all equal. We find that the sum of the charges is\footnote{One sees that this is the same as the analogous result for M5 branes on a disc as in \cite{Bah:2021hei}.}
\be
\sum_{I=1}^{4}Q_I=\frac{1}{k}-\frac{\Delta z}{4\pi}\, .
\ee
Note in particular that the sum of the charges does not equal the Euler characteristic of the disc, and in fact can never be made to. This implies that the mechanism for preserving supersymmetry, like in the spindle case studied above, is not the usual topological twist. One can also see this from computing the explicit Killing spinors on the four-dimensional solution; they depend on the disc coordinates which is not the case for a standard topological twist, see appendix \ref{app:spinors}.


\paragraph{Beyond the $\boldsymbol{\text{T}^3}$ truncation.}

We have argued above that one cannot obtain a disc from our local solution in either the $X^0 X^1$ truncation nor the Einstein--Maxwell truncation. One may wonder if it is possible to have a solution keeping the remaining non-zero $q_I$'s distinct. Studying this more general solution near to the singular point one finds that the uplift interpretation is not simply a smeared M2-brane as in the $T^3$ truncation. It would be interesting to see if it is possible in this case to give the singularity a physical interpretation, we have not been able to rule out a more complicated completion involving a more exotic brane configuration.


\section{Uplift to 11d supergravity}\label{sec:11d}


In this section we discuss the 11d uplift of the 4d AdS$_2$ solutions given in section \ref{sec:4d}. We find that the class of spindle solutions is lifted to a smooth 11d geometry. For the other class we find that the boundary of the topological disc gives rise to a stack of smeared M2-branes, making this singularity of physical nature.

\subsection{STU truncation uplift}

The local solution we considered in section \ref{sec:4d} was originally found in 11d in \cite{Gauntlett:2006ns}. 
The full metric after a little rewriting is given by
\be
\dd s^2=\me^{2 A}\bigg( \dd s^2(\text{AdS}_2) + \frac{f(w)}{P(w)} \dd z^2 +\frac{1}{f(w)} \dd w^2 +\frac{4}{P(w) Y} \sum_{I=1}^4(w-q_{I}) \bigg[ \dd \mu_{I}^2 +\mu_{I}^2 \Big(\dd \phi_{I}-\frac{w}{2(w-q_{I})}\dd z\Big)^2\bigg]\bigg)\, ,\label{11dspindlesol}
\ee
where the $\mu_{I}$'s are the embedding coordinates of an $S^7$ satisfying
\be
\sum_{I=1}^{4} \mu_{I}^2=1\, .
\ee
The functions $f(w)$ and $P(w)$ are as in \eqref{Pandf}, whilst $Y$ and $\me^{A}$ are given by
\begin{align}
Y&=\sum_{I=1}^{4} \mu_{I}^2 (w-q_{I})^{-1}\, ,\nonumber\\
\me^{3A}&= P(w)Y\, .
\end{align}
We see that the constraint on the scalars being positive definite is equivalent to the metric having correct signature in 11d. 

It is also convenient to rewrite the metric in the form of the classification of AdS$_2$ solutions in \cite{Kim:2006qu}. Using the results in \cite{Gauntlett:2006ns} it takes the form
\begin{align}
\dd s^2&=\me^{2 A}\Big(\dd s^2(\text{AdS}_2)+(\dd z+\sigma)^2+\me^{-3 A} \dd s_8^2\Big)\, ,\label{met:class}
\end{align}
with 
\begin{align}
\sigma&=-\frac{2 w}{P(w) Y} \sum_I \mu_I^2 \dd \phi_I\, ,\\
\dd s^2_8&=\frac{P(w)Y}{f(w)}\dd w^2+4 \sum_I(w-q_I)(\dd \mu_I^2+\mu_I^2\dd \phi_I^2)-\frac{4 w^2}{P(w)Y}\bigg(\sum_I \mu_I^2 \dd \phi_I\bigg)^2\, .
\end{align}
In this rewriting the R-symmetry vector is manifest. It is important to note that the R-symmetry vector is 
\be
R_{1\mathrm{d}}=\partial_z\, .
\ee
A change in the gauge of the gauge fields, which can be reabsorbed by a coordinate shift of the $\phi_I$'s will alter as we will see later. Note that the Killing spinors have charge $\frac{1}{2}$ under z. 
The 8d space is K\"ahler with K\"ahler form 
\be
J_8 = 2 \dd w \wedge \sum_I \mu_I^2\, \dd \phi_I +4 \sum_I (w-q_I)\, \mu_I \dd \mu_I  \wedge d\phi_I\, .
\ee

\subsection{General analysis}

We start by computing some general quantities of interest for the family of solutions that we study. We compute the free-energy/entropy of the dual CFT, which for AdS$_2$ solutions is given by 
\be
S=\mathcal{F}=\frac{1}{4 G_2}\, ,
\ee
and we check that the M2-brane fluxes are properly quantized. Furthermore, we check the regularity of the 11d solutions in both the spindle and the disc cases, and examine some interesting regions of the solutions.

\paragraph{Flux quantisation.}

As usual, we need to check that the flux of the solution is quantized so that the solution can be properly lifted to M-theory. The four-form flux of the solution is given by
\begin{equation}
G_4 = L^3 \,\text{dvol}_{\text{AdS}_2} \wedge \big[ \dd \big( e^{3A} (\dd z+\sigma) \big)- J_8 \big]\,.
\end{equation}
There are no non-trivial four-cycles in our geometry, but there are non-trivial seven-cycles, so we focus on the Hodge dual flux which can be written as \cite{Couzens:2018wnk}
\begin{equation}
\ast G_4 = L^6 \, (\dd z + \sigma) \wedge \dd \sigma \wedge \frac{J_8^2}{2} + \text{closed piece} \,.
\end{equation}
Since the closed piece does not contribute to any of the integrals we refrain from presenting it here.

Both the spindle and the topological disc have a seven-cycle given by the $S^7$. In order to compute the flux through this cycle we go to one of the endpoints of the $w$ interval, because there the seven-sphere decouples from the rest of the geometry which allows us to integrate over this cycle. We compute this integral as
\begin{equation}
\frac{1}{(2\pi \ell_p)^6}\int_{\text{S}^7} \ast\, G_4 = \frac{2 L^6}{\pi^2\ell_p^6} \equiv N \,,
\end{equation}
and hence find the quantisation condition $N\in\mathbb{Z}$.

For the spindle this is the only non-trivial seven-cycle. However, for the topological disc we find an additional cycle in the simultaneous $\mu_4, w \to 0$ limit, associated with the presence of smeared M2-branes. We will discuss the quantisation condition arising from this cycle in section \ref{sec:11ddisc}.

\paragraph{Free energy}

In order to compute the free energy, it is convenient to use the metric in the form of \eqref{met:class}.
We find that the two-dimensional Newton's constant is given by the simple expression
\begin{equation}
\begin{aligned}
\frac{1}{G_2} &= \frac{L^9}{G_{11}} \int_{Y_9} e^{9A}\, \text{dvol}_{Y_9} = \frac{8 L^9}{ 3 \pi^3 \ell_p^9}  \, (w_3-w_2) \,\Delta z \,,
\end{aligned}
\end{equation}
where we ahve used that $G_{11} = \frac{(2\pi)^8 \ell_p^9}{16\pi}$, and $\ell_p$ denotes the 11d Planck length. Using the quantisation of the flux and the definition of $N$ above we have
\be
S=\frac{1}{4G_2} = \frac{1}{3\sqrt{2}} N^{3/2}\, (w_3-w_2)\, \Delta z\, .\label{eq:G2gen}
\ee


\subsubsection{Spindle}

Here we study the regularity of the spindle solutions in the uplift to 11d. We will find that the singularities from 4d have disappeared in the uplift, and that the solution is regular. This is a well-known feature of M2- and D3- branes on spindles (though not for M5's), see for example \cite{Ferrero:2020laf,Ferrero:2020twa,Boido:2021szx}, and is sometimes referred to as desingularization.

In 4d we found conical singularities at $w=w_{2}$ and $w=w_{3}$, so here we again focus on these points. Our 11d solution has five U$(1)$ Killing vectors, namely $\partial_z$ and $\partial_{\phi_I}$. In order to check for conical singularities, we are looking for linear combinations of these Killing vectors that have vanishing norm at $w_2$ or $w_3$. It turns out that the only such Killing vectors are given by\footnote{Note that the $z$ circle, that gave rise to the conical singularities in the 4d solution, does not pinch anywhere in the uplifted manifold. The norm of the corresponding Killing vector is simply given by $||\partial_z||^2 = \me^{2A}$.}
\begin{equation}
\partial_{\psi_i} = c_i\left( \partial_z + \sum_I \frac{w_i}{2(w_i-q_I)}\,\partial_{\phi_I} \right) \,,
\end{equation}
where $i=2,3$. For these, we find that $||\partial_{\psi_2}||^2(w_2)=||\partial_{\psi_3}||^2(w_3)=0$. We can now fix the normalization constants $c_i$ by imposing that these Killing vectors have the appropriate periodicity of $2\pi$ around the point where they degenerate. We expand the norm of the $\partial_{\psi_i}$ in $\Delta w = \pm(w-w_i)$ which gives
\begin{equation}
||\partial_{\psi_i}||^2 = \me^{2A}\:\frac{c_i^2\,|f'(w_i)|}{P(w_i)}\,\Delta w + \mathcal{O}\big(\Delta w^2\big) \,,
\end{equation}
and therefore the metric spanned by $w$ and $\psi_i$ close to $w_i$ can be written as
\begin{equation}
\frac{1}{|f'(w_i)|\,\Delta w}\,\dd w^2 + c_i^2\:\frac{|f'(w_i)|}{P(w_i)}\,\Delta w\, \dd \psi_i^2 = \frac{4}{|f'(w_i)|} \left[ \dd R^2 + \frac{c_i^2\,f'(w_i)^2}{4P(w_i)}\,R^2\dd\psi_i^2 \right] \,.
\end{equation}
Here we have omitted the overall warping $\me^{2A}$, and we have made the change of coordinates $\Delta w = R^2$. We see that in order to get the correct periodicity, we must fix the normalization constants as
\begin{equation}
c_i = \left( \frac{f'(w_i)}{2\sqrt{P(w_i)}} \right)^{-1} = \left( \sum_I \frac{w_i}{2(w_i-q_I)} -1 \right)^{-1} \,.
\end{equation}
These coefficients are identical to the ones found in \cite{Gauntlett:2006ns}, where the regularity of this family of solutions was previously studied.

We now have six Killing vectors that degenerate somewhere in the manifold, but there are only five isometries. We therefore know that they must be related by the constraint\footnote{Since it turns out that the constant coefficients of the $\phi_I$ Killing vectors become the magnetic charges $p_I$ defined in \eqref{eq:spindleps}, we have chosen to call them $p_I$ from the outset. Note that there was no assumption that these are magnetic charges in reaching this conclusion.}
\begin{equation}\label{eq:spindledegsum}
a_2\, \partial_{\psi_2}+a_3\, \partial_{\psi_3}+ p_1\, \partial_{\phi_1}+ \ldots + p_4\, \partial_{\phi_4} = 0\,,
\end{equation}
for some coprime integers $a_2, a_3, p_1, \ldots, p_4$. Notice that the constants $c_i$ may be rewritten as
\be
c_2=\frac{n_- \Delta z}{2\pi}\, ,\qquad c_3=-\frac{n_+\Delta z}{2\pi}\, 
\ee 
by using \eqref{eq:spindleperiod}. 
Expanding the terms in the sum \eqref{eq:spindledegsum} we find the constraints
\begin{align}
0&=a_2 n_{-}-a_3 n_+\, ,\\
0&=\sum_{i=2}^{3}\frac{a_i c_i w_i}{2(w_i-q_I)}+ p_I \, .
\end{align}
where the latter must be satisfied for all $I$. We may solve the first by
\be
a_2=n_+\, ,\qquad a_3=n_-\, ,
\ee
where we have used that $a_2$ and $a_3$ should be coprime. Plugging this into the second constraint we find
\be
p_I=n_+ n_- Q_{I}\, ,
\ee
which is precisely the quantisation condition we imposed on the fluxes in \eqref{eq:spindleps}. We conclude that provided the fluxes $p_I$ (defined in \eqref{eq:spindleps}) are relatively prime to both $n_+$ and $n_-$, which are relatively prime themselves, then the 11d solution is smooth and free of conical singularities.

\paragraph{R-symmetry vector.}
Let us now consider the R-symmetry of the solution. Using the general results in \cite{Kim:2006qu,Couzens:2018wnk} and the form presented in \eqref{met:class} we can identify the 1d R-symmetry vector as
\be
R_{1\mathrm{d}}=\partial_z\, .
\ee
However, this leads to a Killing spinor which has charge $\tfrac{1}{2}$ under the isometry of the spindle. 
To see why this is true, note that the 11d Killing spinor is the tensor product of the Killing spinor of the 4d solution with the Killing spinor on $S^7$. The uplift of our solution to 11d, as given in \cite{Gauntlett:2006ns} which uses the gauge fields in \eqref{eq:gauge}, is in the correct form of the classification in \cite{Kim:2006qu} (we follow the conventions in \cite{Couzens:2018wnk} though). This implies that the 11d Killing spinor has charge $\tfrac{1}{2}$ under $z$ and therefore the 4d Killing spinor has the same charge. Now, since the Killing spinor on $S^7$ has charge $\tfrac{1}{2}$ under each of the U$(1)$'s we may absorb the $z$ dependence by a coordinate shift:
\be
\phi_I\rightarrow \tilde{\phi}_I=\phi_I-\frac{1}{4} z\, .
\ee
The R-symmetry vector is then
\be
R_{1\mathrm{d}}=\partial_z+\frac{1}{4}\sum_{I=1}^{4}\partial_{\tilde{\phi}_{I}}\, ,
\ee
and we identify the summand with the 3d superconformal R-symmetry of ABJM before compactification. Note the similarity with the D3 case in \cite{Boido:2021szx}.

\paragraph{Entropy.}
Finally Newton's constant is given in equation \eqref{eq:G2gen},
\be
\frac{1}{G_2} = \frac{2\sqrt{2}}{3} N^{3/2}\, (w_3-w_2)\, \Delta z\, .
\ee
We now want to write this in terms of the charges $Q_I$ given in \eqref{eq:spindleps}. Since the result for the Newton's constant takes the same form in the general STU solution as in the truncated solution we will present the general result for the four charges. To proceed it is useful to note that the term $(w_3-w_2)\Delta z$ appears in the expression for the charges. A simple but tedious computation using the properties of the two polynomials $P(w)$ and $f(w)$ and the four charges allows us to express the roots implicitly in terms of the charges\footnote{See \cite{Couzens:2021cpk} for further details on these expressions in terms of the quartic invariant. }
\begin{align}
\hat{Q}^{(4)}&\equiv\prod_{I=1}^{4} Q_I=\bigg(\frac{\Delta z (w_3-w_2)}{4\pi}\bigg)^4\frac{w_1 w_4}{w_2 w_3}\, ,\\
\hat{Q}^{(3)}&\equiv\sum_{I=1}^4\prod_{J\neq I} Q_J=\bigg(\frac{\Delta z (w_3-w_2)}{4\pi}\bigg)^3\frac{w_1 w_4}{w_2 w_3}\bigg[(w_1+w_4)\frac{w_2 w_3}{w_1 w_4}+(w1+w2-2 w_2-2w_3)\bigg]\, ,\\
\hat{Q}^{(2)}&\equiv\sum_{1\leq I<J\leq4}Q_I Q_J=\bigg(\frac{\Delta z (w_3-w_2)}{4\pi}\bigg)^2\frac{w_1 w_4}{w_2 w_3}\bigg[1+3(w_2+w_3)^2 -2 (w_2+w_3)(w_1+w_4)+w_1 w_4\nonumber\\
&-3 w_2 w_3
+\frac{1}{w_1 w_4}\Big(w_2 w_3+w_2 w_3 \sum_{1\leq I<J\leq 4}w_I w_J+(w_1+w_4-2w_2-2w_3)\sum_{I=1}^{4}\prod_{J\neq I} w_I\Big)\bigg]\, ,\\
\hat{Q}^{(1)}&\equiv\sum_{I=1}^4Q_I=\bigg(\frac{\Delta z (w_3-w_2)}{4\pi}\bigg)\Big(2(w_1+w_4)-(w_2+w_3)-\frac{w_1 w_4}{w_2w_3}(w_2+w_3)\Big)\, .
\end{align}
Note that $Q^{(1)}$ can be expressed in terms of $n_{\pm}$ as in \eqref{eq:spindleQsum} once the constraint on the periods, \eqref{eq:spindleperiod} is taken into account.
To proceed it is useful to define 
\begin{align}
x=\frac{\Delta z(w_3-w_2)}{4\pi}\, ,\qquad  \alpha=w_1+w_4\, ,\qquad \beta=w_1 w_4\, ,
\end{align}
where, as in all good algebra problems, `$x$' is what we want to compute.

We first solve the constraint on the periods, namely equation \eqref{eq:spindleperiod}, for the variable $\beta$ which has solution
\be
\beta=-\frac{w_2 w_3\big(n_-(w_2-\alpha)+n_+(w_3-\alpha)\big)}{n_+ w_2+n_- w_3}\, .
\ee
With this solution we see that both $\hat{Q}^{(1)}$ and $\chi$ take the expected form. We next eliminate $\alpha$ in favour of the new variable $x$ which gives
\be
\alpha=\frac{n_+w_2(1-n_- w_2 x)+n_-w_3(1+n_+ w_3 x)}{n_- n_+ x (w_3-w_2)}\, .
\ee
Substituting the new variables and constraints into the functions of the charges $\hat{Q}^{(A)}$, for $A={2,3,4}$ gives three equations for the three unknowns $x,w_2,w_3$. It is again convenient to change variables, introducing
\be
\gamma=w_3+w_2\, ,\qquad \delta=w_3-w_2\, .
\ee
We may now solve for $x,\gamma,\delta$ in terms of $n_{\pm}$, and $\hat{Q}^{(A)}$, $A={2,3,4}$ giving\footnote{One actually obtains four solutions when inverting these results however the constraints $\delta>0$ and $x>0$ which follow from regularity leave a single physically sensible solution.}
\begin{align}
x&=\sqrt{\frac{1+n_-n_+ \hat{Q}^{(2)}-\sqrt{\big(1+n_- n_+ \hat{Q}_{2}\big)^2-4 n_-^2 n_+^2 \hat{Q}^{(4)}}}{2 n_- n_+}}\, ,\\
\delta&=\frac{(n_-+n_+)x}{\sqrt{\big(1+n_- n_+ \hat{Q}_{2}\big)^2-4 n_-^2 n_+^2 \hat{Q}^{(4)}}}\, ,\\
\gamma&=\frac{\Big((n_+-n_-)\big(1+n_-n_+\hat{Q}^{(2)}\big)+2 n_-^2n_+^2 \hat{Q}^{(3)}\Big)x}{\big(1+n_- n_+ \hat{Q}_{2}\big)^2-4 n_-^2 n_+^2 \hat{Q}^{(4)}}\, .
\end{align}
In principle we can now invert to solve for the roots in terms of the charges however this leads to ugly expressions and so we will spare the reader this pain. Instead, we can now express the entropy in terms of the charges and Euler characteristic as
\be
S=\frac{1}{4G_{2}}=\frac{\pi}{3}N^{\tfrac{3}{2}}\sqrt{4 \hat{Q}^{(2)}+\chi^2-\big(\hat{Q}^{(1)}\big)^{2}-\sqrt{\Big(4 \hat{Q}^{(2)}+\chi^2-\big(\hat{Q}^{(1)}\big)^{2}\Big)^2-64 \hat{Q}^{(4)}}}\, , \label{eq:spindleentropy}
\ee
with $\hat{Q}^{(A)}$ the symmetric combination of the charges defined above. This result is completely general and valid for the full STU solution, not just the $X^0 X^1$ truncated solution we studied in detail in this section.\footnote{In \cite{Ferrero:2021ovq} the authors study solutions in the $X^0 X^1$ truncation and present an expression for the entropy in that truncation. The form given there does not obviously agree with the one presented here when restricted the $X^0 X^1$ case, however one can show that the two expressions give numerically equivalent results. Despite this one can show that this alternative expression for the entropy relies on a relation between the roots $w_I$ which is not true for the full STU solution and therefore cannot be extended to the full multi-charge solution.}


\subsubsection{Topological disc}\label{sec:11ddisc}
\begin{figure}[h!]
\centering
  \includegraphics[width=0.6\linewidth]{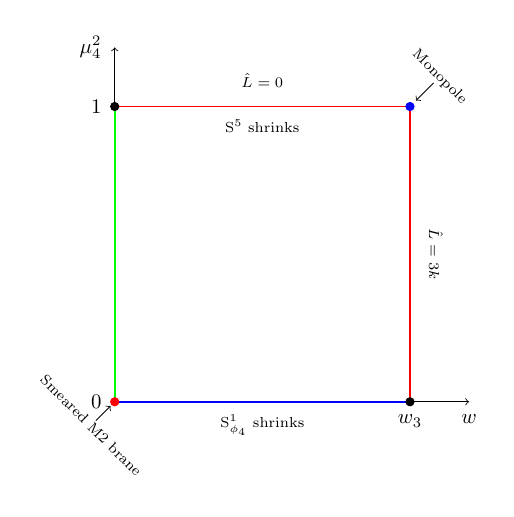}
  \captionsetup{width=.75\linewidth}
  \caption{\textit{A schematic plot of the rectangle over which the} $S^5\times S^1_{\phi_4}\times S^1_z$\textit{ is fibered. The blue line indicates that the $\phi_4$ circle pinches smoothly to $\mathbb{R}^2$. Along the green edge the metric is smooth, but has a singularity consistent with a smeared M2 brane at the intersection with the blue line. At the opposite corner sits a monopole leading to a $\mathbb{R}^8/\mathbb{Z}_k$ orbifold.}}
  \label{fig:disc}
\end{figure}

In this section we will study the different regimes of the topological disc solution. The internal metric takes the form of an $S^5\times S^1_{\phi_4}\times S^{1}_z$ fibration over the rectangle given by $(\mu_4^2,w)$ with $\mu_4^2\in[0,1]$ and $w\in [0,w_3]$. Within the rectangle the manifold does not degenerate, but on each of the boundaries some part of the metric degenerates, see figure \ref{fig:disc} for a pictorial representation of the solution.  

Before we begin it is convenient to reparametrise the $S^7$ embedding coordinates as 
\be
\mu_i=\sqrt{1-\mu_4^2}\,m_i\, ,  \qquad \text{with}\quad \sum_{i=1}^3 m_i^2=1\, ,
\ee
and to define
\be
Y=\frac{1}{w}\Big(\mu_4^2+ w(1-\mu_4^2)\,\hat{Y}\Big)\, ,\qquad \hat{Y}=\sum_{i}\frac{m_i^2}{w-q_i}\, ,\quad  P(w)=w \hat{P}(w)\, ,\quad f(w)=w \hat{f}(w)\, .
\ee
Note that $w Y$ is non-zero except at $w=\mu_4=0$ and $\hat{Y}$ is strictly positive definite. 
With these definitions the metric takes the form
\begin{align}
\dd s^2=&\Big[\hat{P}(w)\Big(\mu_4^2+ w (1-\mu_4^2)\, \hat{Y}\Big)\Big]^{2/3}\Bigg(\dd s^2 (\text{AdS}_2)+\frac{\hat{f}(w)}{\hat{P}(w)}\dd z^2 +\frac{1}{w \hat{f}(w)}\dd w^2 \nonumber\\
&+\frac{4(1-\mu_4^2)}{\hat{P}(w)\Big(\mu_4^2+w (1-\mu_4^2)\, \hat{Y}\Big)}\sum_{i=1}^{3}(w-q_i)\Big[\dd  m_i^2+m_i^2 D\phi_i^2\Big]\\
&+\frac{4}{\hat{P}(w)\Big(\mu_4^2+w (1-\mu_4^2)\, \hat{Y}\Big)}\Big[\dd \mu_4^2+\mu_4^2 D \phi_4^2+ \sum_{i=1}^{3}(w-q_i)\Big( \frac{m_i^2\mu_4^2 }{1-\mu_4^2} \dd \mu_4^2-2m_i\mu_4 \dd \mu_4 \dd m_i \Big)\bigg]\Bigg)\nonumber\, .
\end{align}
Note that the cross terms $\dd \mu_4\dd m_i$ vanish if all $q_i$ are set equal.

\paragraph{Smeared M2-branes.}
Let us begin with the line $\mu_4^2=1$ away from the end-points of $w$. We see that this point corresponds to the degeneration of the (squashed) $S^5$, with both the $S^1_z$ and $S^{1}_{\phi_4}$ circles remaining of finite size. However, it is only for $q_1=q_2=q_3\equiv q$ that the degeneration is smooth and gives $\mathbb{R}^6$. We will therefore restrict to the T$^3$ truncation from now on.\footnote{Note that this simplifies the metric considerably. In particular we have $\hat{Y}=(w-q)^{-1}$ and the cross terms $\dd\mu_4 \dd m_i$ drop out.} Conversely, consider the degeneration at $\mu_4=0$ away from the end-points of $w$. We see that the only part of the metric which degenerates is the circle $S^1_{\phi_4}$, degenerating smoothly if $\phi_4$ has period $2\pi$. This is of course the required period for the round $S^7$ in the uplift. 

Next let us consider the degeneration at $w=0$ away from both end-points of $\mu_4$. Since all the hatted objects, $\hat{P}(w), \hat{f}(w)$ and $\hat{Y}$ are non-zero and positive at $w=0$ the only degeneration of the metric is the line-interval and the metric is therefore smooth. In particular, note that the singularity that we encountered at $w=0$ in 4d has been removed in the uplift. Instead, we have a singularity at $w=\mu_4=0$. To investigate this properly it is convenient to make the change of coordinates
\be\label{flavour_M2_parametr}
\hat{P}(0) w= -q  r\cos^2 \bigg( \frac{\theta}{2}\bigg)\, ,\quad \hat{P}(0) \mu_4^2=  r \sin^2 \bigg(\frac{\theta}{2}\bigg)\, ,
\ee
and then expand around $r=0$, which gives the metric in the form
\begin{align}
\dd s^2 =r^{2/3}\bigg( \dd s^2 (\text{AdS}_2) +\dd z^2 \bigg)+\frac{|q|}{r^{1/3}}\bigg(\dd r^2 +r^2 \Big(\dd\theta^2+\sin^2 \theta \dd \phi_4^2\Big) + \sum_{i} \Big[\dd m_i^2 +m_i^2 D \phi_i^2\Big]\bigg)\, .
\end{align}
In comparing with appendix \ref{app:smeared}, where we study the metric of various smeared M2-branes, we see that this takes the form of an M2-brane with world-volume AdS$_2\times S^{1}_z$, localised at the centre of $\mathbb{R}^3$ and smeared on a round $S^5$. The singularity is meaningful, it arises due to a flavour M2-brane in the geometry. Similar results were found in the M5 \cite{Bah:2021hei,Suh:2021ifj}, D3 \cite{Couzens:2021tnv, Suh:2021ifj} and D4-D8 system \cite{Suh:2021aik} cases. Note that we could have taken the limits separately, and in either order, and obtained the same result, the change of coordinates was introduced for convenience. It is also simple to see that the limit $w=0, \mu_4=1$ is smooth given the discussion above. It is interesting to note that the location of the smeared brane for both D3- and M2-branes is the bottom most corner of the diagram in figure \ref{fig:disc}, for the M5-branes it is located along the $w=0$ side \cite{Bah:2021hei}. This seems to be a general feature of odd dimensional spheres, the smeared brane is located at a single point in the uplift, whilst for even dimensional spheres it is located along a line. One also sees a different behaviour between the uplift of spindle solutions on even and odd-dimensional spheres. Uplifting on odd-dimensional spheres allows for the orbifold singularities to be resolved whilst on even-dimensional spheres orbifold singularities remain in the uplifted solution. 

\paragraph{Monopoles.}
We have left the most subtle limits for last. We first want to consider the $w=w_3$ limit before taking the simultaneous $w=w_3$ $\mu_4^2=1$ limit. For this it is best to rewrite the metric as a fibration of the $S^1_z$ over the $S^7$. In doing this rewriting it is convenient to allow an arbitrary gauge choice for the four gauge fields of the form $\delta A_I=n_I \dd z$. There are two convenient choices that one can make. The first is such that the gauge field vanishes at the orbifold point, that is we take 
\be
\delta A_{I}=\frac{w_3}{2(w_3-q_I)}\dd z\, .
\ee
This choice of gauge has the benefit of leading to a well defined gauge field in 4d when going to the orbifold point, recall that $z$ shrinks there, however it leads to a Killing spinor which is charged under the isometry of the disc. Following our earlier discussion above we may perform a gauge transformation of the gauge fields so that the spinor is uncharged under the rotation generator of the disc. It turns out that this is the best choice of gauge to make.

First note that $A_4$ is pure gauge at the moment.  We should therefore first perform a gauge transformation to set $A_4=0$, that is we have $A_4\rightarrow A_4+\frac{1}{2}\dd z=0$. This is equivalent to performing a change of coordinates $\phi_4\rightarrow\tilde{\phi}_4= \phi_4-\frac{1}{2}z$, since this preserves the form of $\dd\phi_4+A_4$. 
Now consider the phase of the Killing spinor after this transformation: it is proportional to 
\be
\frac{z}{2}+\sum_{i=1}^{3}\phi_i+\tilde{\phi}_4\, .
\ee
We may now perform the coordinate shift $\phi_i\rightarrow \tilde{\phi}_{i}=\phi_i-\frac{1}{6} z$ which removes the $z$ dependence of the phase of the Killing spinor. 
We can ensure that the one-forms $\dd \phi_i+A_i$ are invariant under this coordinate transformation provided we perform a gauge transformation 
\begin{equation}
A_i\rightarrow A_i+\frac{1}{6} \dd z\, ,\qquad A_4\rightarrow A_4+\frac{1}{2}\dd z\, .
\end{equation}
After this slightly round-about argument we conclude that the gauge transformation above is equivalent to removing the $z$ dependence in the 4d Killing spinor and the expressions above we should set
\be
n_i=\tfrac{1}{6} \qquad \text{and}\qquad  n_4=\tfrac{1}{2} \, .
\ee
Notice that the gauge choice we make here will not change the analysis that we have performed earlier in this section, however it has an important effect here. 

Rewriting the metric in the form of an $S^1_z$ fibration over the seven sphere we have
\begin{align}
\dd s^2&=\me^{2 A}\bigg[ \dd s^2(\text{AdS}_2)+\frac{1}{w \hat{f}(w)}\dd w^2+ \frac{4(w-q)(1-\mu_4^2)}{\hat{P}(w)(w-q \mu_4^2)}\sum_{i=1}^{3}\dd m_i^2+ R_z \Big(\dd z-L \sum_{i=1}^{3}m_i^2 \dd \phi_{i}\Big)^2\\
&+R_1m_1^2 \Big(\dd \phi_1- S_1\sum_{i=2}^{3}m_i^2 \dd \phi_i\Big)^2+R_2 m_2^2\Big(\dd \phi_2-S_2 m_3^2 \dd \phi_3\Big)^2+R_3m_3^2\dd \phi_3^2+\frac{4 w (w-q)\mu_4^2}{\hat{P}(w)(w-q \mu_4^2)}\dd \phi_4^2
\bigg]\, , \nonumber
\end{align}
where
\begin{align}
R_z&=\frac{(q+2w)^2(1-\mu_4^2)+9(w-q\mu_4^2)\hat{f}(w)}{9\hat{P}(w)}\, ,\nonumber\\
R_1&=\frac{4(w-q)^2 (1-\mu_4^2)\Big((1-m_1^2)(2w+q)^2 (1-\mu_4^2)+9(w-q \mu_4^2)\hat{f}(w)\Big)}{(w-q\mu_4^2)\hat{P}(w)\Big((2w+q)^2 (1-\mu_4^2)+9 (w-q \mu_4^2)\hat{f}(w)\Big)}\, ,\nonumber\\
R_2&=\frac{4(w-q)^2(1-\mu_4^2)\Big(m_3^2 (2w+q)^2(1-\mu_4^2)+9 (w-q\mu_4^2)\hat{f}(w)\Big)}{(w-q \mu_4^2)\hat{P}(w)\Big((1-m_1^2)(2w+q)^2(1-\mu_4^2)+9(w-q \mu_4^2)\hat{f}(w)\Big)}\, ,\nonumber\\
R_3&=\frac{36(w-q)^2 (1-\mu_4^2)\hat{f}(w)}{\hat{P}(w)\Big(m_3^2 (2w+q)^2 (1-\mu_4^2)+9(w-q\mu_4^2) \hat{f}(w)\Big)}\, ,\\
L&=\frac{6(w-q)(2w+q)(1-\mu_4^2)}{(2w+q)^2(1-\mu_4^2)+9(w-q \mu_4^2)\hat{f}(w)}\, ,\nonumber\\
S_1&=\frac{(2w+q)^2 (1-\mu_4^2)}{(1-m_1^2)(2w+q)^2 (1-\mu_4^2)+9 (w-q\mu_4^2)\hat{f}(w)}\, ,\nonumber\\
S_2&=\frac{(2w+q)^2 (1-\mu_4^2)}{m_3^2 (2w+q)^2(1-\mu_4^2)+9(w-q\mu_4^2)\hat{f}(w)}\, .\nonumber
\end{align}

From the form of the functions it is clear that all the $R_i$ vanish as $\mu_4^2\rightarrow 1$, whilst $R_z$ remains finite. In addition the fibration functions $S_{1,2}$ vanish in this limit and we have
\be
L(\mu_4^2=1,w)=0\, .
\ee
This leads to the smooth shrinking of the $S^5$ that we saw previously. 
Conversely, at $w=w_3$ the $R_i$ are all finite, $S_{1,2}$ become constant (in particular the $\mu_4$ dependence drops out) and the $S^5$ has a non-zero radius but is twisted. Note that the function $L$ is a non-zero constant at $w=w_3$,
\be
L(\mu_4^2, w=w_3)=\frac{6(w_3-q)}{2 w_3+q}= -\frac{3 k \Delta z}{2\pi}\equiv\frac{\hat{L}\Delta z}{2\pi}\, .
\ee
Clearly there is a jump at the corner $(\mu_4^2=1, w=w_3)$ of the rectangle depending on the direction we approach the corner from. This signifies the existence of a monopole source located there. The charge of the monopole is computed by evaluating the Chern number of the line-bundle, 
\be
Q_m=\frac{1}{\Delta z}\int\dd Dz\, .
\ee 
For the case at hand we find that the monopole charge is 
\be
Q_m^{i}=\hat{L}=3 k\, ,
\ee
and this accounts for the singularity at the origin of the disc in 4d.

Since this is somewhat subtle let us study this in a different way. Lets take the simultaneous $\mu_4^2=1, w=w_3$ limit by changing to the coordinates
\be
\mu_4^2=1-r^2 \sin^2\xi\, ,\qquad w=w_3+{(w_3-q) \hat{f}'(w_3)} r^2 \cos^2\xi\, .
\ee
In this limit the metric becomes
\begin{align}
\dd s^2&=w_3^{2/3}\bigg[\dd s^2(\text{AdS}_2)+4 \dd \phi_4^2 \\
&+\frac{4 }{w_3^{2/3}}\bigg\{\dd r^2+ r^2 \bigg(\dd \xi^2+ \frac{\hat{f}'(w_3)^2}{4}\cos^2 \xi \dd z^2+\sin^2 \xi  \sum_{i=1}^{3}\Big[\dd m_i^2+m_i^2   \Big(\dd \phi_i+ \frac{\hat{f}'(w_3)}{6}\dd z\Big)^2\Big]\bigg)\bigg\}\bigg]\, .\nonumber
\end{align}
This is $\text{AdS}_2\times S^1_{\phi_4}\times \mathbb{R}^8/\mathbb{Z}_k$ where we have used
\eqref{Deltazdisc}, and we see that the orbifold singularity at the centre of the disc in 4d arises in 11d from a quotient space $\mathbb{R}^8/\mathbb{Z}_k$, i.e. it is a monopole. A natural interpretation is that this corresponds to a regular puncture whilst the smeared M2 brane arises due to an irregular puncture. It would be interesting to understand this from a field theory computation.

\paragraph{R-symmetry vector.}
We conclude the regularity analysis by identifying the R-symmetry of the solution. From the general form of the metric without the gauge shifts the R-symmetry vector is simply
\be
R_{1\mathrm{d}}=\partial_z\, .
\ee
However, recall that we performed a gauge transformation above in order for the Killing spinor to be independent of the spindle U$(1)$ coordinate. Taking into account the gauge transformation and denoting the new coordinates $\tilde{\phi}_I$ we have
\be
R_{1\text{d}}=\partial_z+\frac{1}{6}\sum_{i=1}^3\partial_{\tilde{\phi}_i}-\frac{1}{2}\partial_{\tilde{\phi}_4}\, .\label{eq:Rsymdisc}
\ee
Note that this is different to the result of the spindle. The $\tilde{\phi}$ terms do not give the canonical R-symmetry for the parent ABJM theory as in the spindle example.

\paragraph{Flux quantisation.}

As mentioned earlier, there are two more cycles that we should consider when quantising the flux. The first is the $S^7/\mathbb{Z}_k$ lying at the top right rectangle in figure \ref{fig:disc} where the monopole is located. We find that the quantisation of the flux imposes that $N/k$ is integer. We shall therefore define 
\be
N=k \hat{N}\, ,\qquad \hat{N}\in \mathbb{Z}.\label{eq:kN}
\ee
The second extra cycle we will consider is the one located at the left-most corner of the rectangle, that is the limiting point where the smeared M2 brane is located. This cycle is given by $S^5 \times S^1_z \times I_\theta$, using the parametrisation \eqref{flavour_M2_parametr}. Integration of $\ast\, G_4$ over this cycle yields the quantisation condition
\begin{equation}\label{flavour_flux}
\frac{1}{(2\pi \ell_p)^6}\int_{S^5 \times S^1_z \times I_\theta} \ast \,G_4 = \frac{L^6}{\pi^2 \ell_p^6}\,   \frac{\Delta z}{2\pi}= \frac{N}{2}  \frac{\Delta z}{2\pi} \in\mathbb{Z}\,.
\end{equation}
We may rewrite this in terms of the charges which gives
\be
N \bigg(\frac{1}{k}+3 Q\bigg)\in \mathbb{Z}\, .
\ee
We see that this is generically satisfied once $N$ is defined as in \eqref{eq:kN}.

\paragraph{Entropy.}
We may now compute the 2d Newton's constant and thereby the entropy. We find
\be
S=\frac{1}{4G_2}= \frac{2\sqrt{2}\pi N^{3/2}}{3k} \sqrt{\frac{p^3}{1+3p}}\, ,
\ee
with $Q=p k$. Contrast this expression with the entropy for the spindle in \eqref{eq:spindleentropy}.

\section{Conclusions}\label{sec:conclusion}

In this work we have considered holographically, compactifications of M2 branes on Riemann surfaces with non-constant curvature. We have shown that the disc and spindle solutions can be obtained from different global completions of the same local solution and analysed their properties in detail. One of the key features of these solutions is a that the Killing spinor is a section of a non-trivial bundle over the compactification Riemann surface in contrast to the more standard topological twist. 

It would be interesting to consider the dual field theory computation computing the  index of ABJM on the spindle and topological disc. In contrast to the usual topological twist solutions, \cite{Benini:2015eyy} one must twist the theory with the isometry of the Riemann surface. For the D3- and M5-branes cases the anomaly polynomial gave an efficient method of doing this, however such a computation is not possible for the theory studied here. 

Another interesting direction is to consider the spinning generalisation of the solutions studied here, see for example \cite{Faedo:2021kur,Ferrero:2021ovq}. Since the non-rotating Einstein--Maxwell solutions studied in \cite{Ferrero:2020twa} are a sub-class of the solutions studied here, one would expect that such a rotating solution is also possible for the multi-charge solutions with four charges. Such solutions would fit into the classification of \cite{Couzens:2020jgx} which studied the near-horizon of rotating black holes in M-theory. This would also allow for the possibility of spinning discs following the ideas presented in this paper on the different choices of global completions of the local solutions. Such a generalisation has been studied for rotating multi-charge D3-branes in \cite{Hosseini:2021fge} and it would be interesting to study the embedding of those solutions in the classification in \cite{Couzens:2020jgx}. Understanding how \cite{Couzens:2020jgx} seems to encompass both magnetic AdS and Kerr--Newmann AdS solutions is also of particular interest and may be used to generate new rotating solutions. 

In \cite{Kim:2020qec}, it was shown that the dyonic black holes of \cite{Halmagyi:2013sla} when uplifted to 11d contain a non-trivial transgression term for the four-form flux. It would be interesting to study if this additional magnetic charge is compatible with the spindle and topological disc solutions studied here and what it corresponds to. Research in these directions is currently underway. 


\section*{Acknowledgements}

We would like to thank Eric Marcus for useful discussions and collaboration on \cite{Couzens:2020jgx}.
CC thanks KIAS for their hospitality during the latter stages of this work and Niall Macpherson and Achilleas Passias for collaboration on \cite{Couzens:2021tnv}. CC is supported by the National Research Foundation of Korea (NRF) grant 2019R1A2C2004880.


\appendix

\section{Smeared M2 branes}\label{app:smeared}

In this appendix we will study the form of the metric for a single stack of M2-branes smeared over internal manifolds of various dimensions, see \cite{Lozano:2019emq} for D-branes of this form. Our interest is in the singularity structure of the solutions, so that we may compare with the singularities we find in the topological disc solutions of section \ref{sec:11ddisc}. 

To begin, recall that the metric of a single stack of M2 branes in flat space, takes the form
\be
\dd s^2=H^{-2/3} \dd s^2(\M^{1,2})+ H^{1/3}\dd s^{2}_{8d}\, .\label{M2metric}
\ee
The M2-branes lie along $\M^{1,2}$ whilst the function $H$ is harmonic on the transverse 8d space. It is typical to impose that the transverse space takes the form of a cone with base a compact manifold and that the function $H$ depends only on the radial coordinate. Indeed writing the transverse space as a cone
\be
\dd s^2_{8d}= \dd r^2 +r^2\dd s^2(X_7)\, ,
\ee 
and taking $H$ to be a function of the radial coordinate only, the harmonic function is given by 
\be 
H(r)=\alpha+\frac{\beta}{r^6}\,,
\ee
with $\alpha$ and $\beta$ two integration constants.
This solution describes a stack of M2-branes localised at the tip of the cone, however this is not the most general solution one can construct. 

Rather than considering a cone, one may decompose the space into the direct product of two pieces of dimensions $s$ and $8-s$ respectively. The first part is a space over which we will smear the M2 brane, whilst the second will be taken to be a cone. Concretely we decompose the transverse 8d space as
\be
\dd s^2_{8d}=\dd r^2 + r^2 \dd s^{2}(X_{7-s})+\dd s^{2}(Y_{s})\, ,
\ee
and impose that the function $H$ is depends only on the radial coordinate once again. 
The function $H$ now takes the form
\be
H=\alpha+\frac{\beta}{r^{6-s}}\, .
\ee
For $s=0$ this clearly reduces to the usual M2 brane solution. 

We may now insert the expressions for the harmonic function into the full brane solution and study the solution close to the brane in the $r\rightarrow 0$ limit. 
Close to $r=0$ the fully localized M2 brane metric looks like
\be
\dd s^2\sim r^4 \dd s^2(\M^{1,2})+r^{-2} \Big(\dd r^2+ \dd s^2 (X_7)\Big)\, ,
\ee
whilst the smeared metric looks like
\be
\dd s^2\sim r^{4-2s/3}\dd s^{2}(\M^{1,2})+r^{s/3-2}\Big(\dd r^2+ r^{2}\dd s^{2}(X_{7-s})+\dd s^2 (Y_s)\Big)\,.
\ee
We will use this result to show that the naively singular solution obtained in section \ref{sec:11ddisc} is in fact indicating the presence of a smeared M2 brane.

\section{Killing spinors of the multi-charge solution}\label{app:spinors}

In this appendix we give the Killing spinors of the 4d AdS$_2\times \Sigma$ solution. In order to use a consistent set of conventions for the supersymmetry transformations and the equations of motion that our solution solve we will first review how the action \eqref{eq:action} is embedded into the general classification of 4d $\mathcal{N}=2$ gauged supergravity in the presence of vector multiplets. We will mostly follow the conventions in \cite{Lauria:2020rhc} for 4d $\mathcal{N}=2$ gauged supergravity. The bosonic part of the Lagrangian is\footnote{Note that we redefine the fields in \cite{Lauria:2020rhc} in order for the Newton's constant to be an overall prefactor.}
\begin{align}
16 \pi G_N\mathcal{L}= \Big(R - 2 g_{\alpha \bar{\beta}}\partial_{\mu}z^{\alpha} \partial^{\mu}\bar{z}^{\bar{\beta}}-2\mathcal{V}\Big)\star 1+ \text{Im}[\mathcal{N}_{IJ}] F^{I}\wedge \star F^{J} -\text{Re}[\mathcal{N}_{IJ}]F^{I}\wedge F^J\, ,
\end{align}
where
\begin{align}
\me^{-\mathcal{K}}&=\ii \big(X^{(I)} \bar{\mathcal{F}}_{I}-\bar{X}^{(I)}\mathcal{F}_{I}\big)\, ,\\
g_{\alpha\bar{\beta}}&=\partial_\alpha\partial_{\bar{\beta}}\,\mathcal{K}\, ,\\
\mathcal{N}_{IJ}&=\bar{\mathcal{F}}_{IJ}+\ii \frac{N_{IK}N_{JL}X^{(K)}X^{(L)}}{N_{PQ}X^{(P)}X^{(Q)}}\, ,\\
N_{IJ}&=2 \Im \mathcal{F}_{IJ}\, ,\quad\mathcal{F}_{I}=\partial_{I} F\, , \quad \mathcal{F}_{IJ}=\partial_I \partial_J F\, ,\\
\mathcal{V}&=-3 X^{(I)}\bar{X}^{(J)}\vec{P}_{I}\cdot \vec{P}_{J}+g^{\alpha\bar{\beta}}\nabla_{\alpha}\nabla_{\bar{\beta}}\bar{X}^{(J)}\vec{P}_{I}\cdot \vec{P}_{J}
\end{align}
and $F$ is the prepotential. The $\vec{P}_{I}$ are moment maps which are taken to be constant here and will be aligned as
\be
P_{I}^{r}=\delta^{3r}\zeta_{I}\, .
\ee
For the prepotential 
\be
F=-\ii \sqrt{X^{(1)}X^{(2)}X^{(3)}X^{(4)}}\, ,
\ee
this agrees with the Lagrangian in \eqref{eq:action}. It is convenient to introduce the three independent scalars (we set axions to vanish in the following) via
\be
\frac{X^{(1)}}{X^{(4)}}=\tau_2\tau_3 \, ,\quad \frac{X^{(2)}}{X^{(4)}}=\tau_1\tau_3\, ,\quad \frac{X^{(3)}}{X^{(4)}}=\tau_1\tau_2\, ,\quad \tau_{i}=\me^{-\phi_i}\, .
\ee
The Lagrangian presented in the main text was understood to be subject to the gauge condition 
\be
X^{(1)}X^{(2)}X^{(3)}X^{(4)}=1\, .
\ee

The gravitino supersymmetry variation is 
\begin{align}
\delta \psi^{i}_{\mu}&=\Big(\nabla_{\mu}-\frac{\ii}{2} \mathcal{A}_{\mu}\Big)\epsilon^{i}-\frac{\ii}{2}\zeta_IA^{I}_{\mu} \tensor{\sigma}{^{3}_{j}^{i}}\epsilon^{j}+\me^{\mathcal{K}/2}X^{(I)}\Im[\mathcal{N}_{IJ}]\slashed{F}^{J}\epsilon^{ij}\gamma_{\mu}\epsilon_{j}+\frac{\ii}{2}\me^{\mathcal{K}/2}\zeta_{I}X^{(I)}\sigma^{3\, ij}\gamma_{\mu}\epsilon_{j}\, ,
\end{align}
whilst for the gaugino it is
\be
\delta \chi_{i}^{\alpha}=\slashed{\partial} z^{\alpha}\epsilon_{i}+\frac{1}{4}g^{\alpha\bar{\beta}}\nabla_{\bar{\beta}}\bar{X}^{(I)} N_{IJ}\big( F^{J}-\ii \star F^{J})_{ab}\gamma^{ab}\epsilon_{ij}\epsilon^{j}- \ii g^{\alpha\bar{\beta}}\nabla_{\bar{\beta}}\bar{X}^{(I)}\tensor{\sigma}{^{3}_{ij}}\epsilon^{j}\, .
\ee
Spinors are taken to be chiral with raised indices and anti-chiral for lowered indices, that is
\be
\gamma_5\epsilon^{i}=\epsilon^{i}\, ,\qquad \gamma_{5}\epsilon_{i}=-\epsilon_i\, .
\ee
The supersymmetry parameters are taken to be Majorana, with charge conjugation defined as
\be
\epsilon^{i}=\epsilon_{i}^{C}\, .
\ee
Following \cite{Cacciatori:2008ek} we combine the Majorana spinors into a Dirac spinor, $\psi\equiv \psi^1+\psi_2$ and $\epsilon=\epsilon^1+\epsilon_2$ which leads to the supersymmetry variations
\begin{align}
\delta \psi=&\bigg[\nabla_{\mu}-\frac{\ii}{2}\mathcal{A}_{\mu}\gamma_5+\frac{\ii}{2}\zeta_{I}A^{I}_{\mu} +\me^{\mathcal{K}/2}\Im\mathcal{N}_{IJ}\slashed{F}^{J}\gamma_{\mu}\Big(\Re X^{(I)}-\ii \Im X^{(I)}\gamma_5\Big)\nonumber\\
&-\frac{\ii}{2}\gamma_{\mu}\zeta_{I}\Big(\Re X^{(I)}-\ii \Im X^{(I)}\gamma_5\Big)\bigg]\epsilon\, .\label{eq:gravKSE}
\end{align}
In the following we will not need the gravitini Killing spinor equation in this form so we suppress the details. 

In order to solve the Killing spinor equations we will take the following basis of 4d gamma matrices,
\be
\gamma_{0}=\ii \sigma_2\otimes \sigma_3\, ,\quad \gamma_1=\sigma_3\otimes \sigma_3\, ,\quad \gamma_2=1_{2\times2}\otimes \sigma_1\, ,\quad \gamma_3=1_{2\times2}\otimes\sigma_2\, ,
\ee
and take the unit radius metric on AdS$_2$ to be
\be
\dd s^2(\text{AdS}_2)=-r^2\dd t^2+\frac{\dd r^2}{r^2}\, .\, 
\ee
The Killing spinors on AdS$_2$, $\eta$ satisfy the Killing spinor equation
\be
\Big[\hat{\nabla}_a-\frac{ \sigma}{2}\rho_a\Big]\eta=0\, ,
\ee
with $\sigma$ a sign that we will determine later, and $\rho_{a}$ the 2d gamma matrices which can be obtained from the above 4d gamma matrices. Note that if $\eta$ solves the Killing spinor equation for $\sigma$ then $\rho_3\eta$ solves the Killing spinor equation for $-\sigma$. It is simple to show that the solutions to the Killing spinor equations are
\be
\eta_{+}=\Big(a_2 r^{-\tfrac{1}{2}} , \sqrt{r}(a_1+a_2 t)\Big)\, ,\qquad \eta_-=\Big(\sqrt{r}(a_1+a_2 t),a_2 r^{-\tfrac{1}{2}} \Big)\, .
\ee

We can decompose the 4d spinors, depending on the sign $\sigma$, in terms of the tensor product of the spinor on AdS$_2$ and the spinor on $\Sigma$, whether it be the spindle or disc. We have
\be
\epsilon_{\pm}=\eta_{\pm}\otimes \theta_{\pm}\, ,
\ee
and the two-component spinor $\theta_{\pm}$ on $\Sigma$ satisfies the projection condition
\be
\sigma_3\theta_{\pm}=\pm \theta_{\pm}\, .
\ee
We may now insert this spinor ansatz into the supersymmetry condition \eqref{eq:gravKSE}. Reducing on AdS$_2$ we find
\begin{align}
\delta\Psi_{a}&= \bigg[\hat{\nabla}_{a}-\frac{\ii \sigma}{2}\gamma_{23}\gamma_{a}+\frac{P'(w)}{4 \sqrt{P(w)}}\Big(\frac{\sigma}{2}-\frac{\sqrt{f(w)}}{2\sqrt{P(w)}}\gamma_{2}+\frac{\ii \sigma w}{2\sqrt{P(w)}}\gamma_{23}\Big)\gamma_a\bigg]\epsilon\, ,
\end{align}
where we have introduced the sign $\alpha$ which satisfies $w-q_I=\sigma|w-q_I|$. The large bracketed term is a projection condition that we must impose on the spinors $\theta_{\pm}$. The resultant Killing spinor equation becomes
\be
\Big[\hat{\nabla}_a-\frac{\ii \sigma}{2}\gamma_{a}\gamma_{23}\Big]\eta_{\pm}\otimes\theta_{\pm}=0\, ,
\ee
which is immediately satisfied by our decomposition if we take $\eta_+$ for $\sigma=1$ and $\eta_-$ for $\sigma=-1$. It remains to solve the two remaining components of the gravitino Killing spinor equation. We introduce an arbitrary gauge shift for the gauge fields of the form
\be
A^I\rightarrow A^I+n^I\dd z\, ,
\ee
which is the same gauge we introduced to study the R-symmetry vector in the main text. Inserting the ansatz into the Killing spinor equation we find the solution
\be
\theta_+=P(w)^{-1/8}\me^{\tfrac{\ii z}{2}\big(1-\sum_I n^I\big)}\bigg(\sqrt{\sqrt{P(w)}+w}\, ,~-\sqrt{\sqrt{P(w)}-w}\,\bigg)\, ,
\ee
and 
\be
\theta_-=\sigma_3\cdot\theta_+=P(w)^{-1/8}\me^{\tfrac{\ii z}{2}\big(1-\sum_I n^I\big)}\bigg(\sqrt{\sqrt{P(w)}+w}\, ,~\sqrt{\sqrt{P(w)}-w}\,\bigg)\, .
\ee
One can then check that this solves the gravitini Killing spinor equations. Note that for the spindle we may remove the phase by taking $n^I=\tfrac{1}{4}$ as claimed in the main text. For the disc, note that the gauge transformation for $A^4$ is further shifted as discussed in section \ref{sec:11ddisc}. Taking this into account we indeed find the R-symmetry vector given in \eqref{eq:Rsymdisc}.
We see that the Killing spinors are manifestly non-constant and therefore we see immediately that supersymmetry on either the spindle or the disc is not preserved by the usual topological twist. 

First let us focus on the spindle case. Note that we have the identity
\be
\sqrt{f(w)}=\sqrt{\sqrt{P(w)}+w}\, \cdot\sqrt{\sqrt{P(w)}-w}\, ,
\ee
moreover at a root of $f(w)$ we have
\be
P(w_*)=w_*^2\, .
\ee
We therefore see that the Killing spinor is never vanishing. Only at the poles of the spindle does a component of the spinor vanish and for our solutions which preserve supersymmetry via the anti-twist, where one root is positive and the second is negative, different components of the spinor vanish. At the poles of the Killing spinor the preserved Killing spinor become chiral and for the anti-twist are of different chiralities on the two halves of the spindle. 

The disc is slightly more subtle. At the non-zero root we have the same discussion as for the spindle at the positive root. We obtain a chiral spinor. At $w=0$ the spinor actually vanishes as $w^{1/8}$. Dividing through by this vanishing conformal factor the resultant spinor is the sum of a chiral and anti-chiral spinor. 


\bibliographystyle{JHEP}

\bibliography{ADSCFT}

\end{document}